\documentstyle[epsfig]{mn2e}

\begin{document}

\def\beq{\begin{equation}}
\def\eeq{\end{equation}}
\def\bea{\begin{eqnarray}}
\def\eea{\end{eqnarray}}
\def\dd{\partial}
\def\LL{{\cal L}}
\def\Om{{\Omega_m}}
\def\om{{\omega_m}}
\def\equ{{\mathrm equ}}

\title[Optimizing BAO Surveys for $\Lambda$CDM]{Optimizing
  Baryon Acoustic Oscillation Surveys -- \\ I:  Testing the 
  concordance $\Lambda$CDM cosmology}

\author[David Parkinson et al.]{\parbox[t]{\textwidth}{David
    Parkinson\thanks{drp21@sussex.ac.uk}$^1$, Chris Blake$^{2,3}$, Martin Kunz$^4$, Bruce
    A. Bassett$^{5,6,7}$, Robert C. Nichol$^5$, and Karl
    Glazebrook$^3$} \\ \\ $^1$ Astronomy Centre, University of Sussex,
  Brighton, BN1 9QH, U.K. \\ $^2$ Department of Physics \& Astronomy,
  University of British Columbia, 6224 Agricultural Road, Vancouver,
  B.C., V6T 1Z1, Canada \\ $^3$ Centre for Astrophysics \&
  Supercomputing, Swinburne University of Technology, P.O.Box 218,
  Hawthorn, VIC 3122, Australia \\ $^4$ Department of Theoretical
  Physics, University of Geneva, 24 quai Ernest Ansermet, CH-1211
  Geneva 4, Switzerland \\ $^5$ Institute of Cosmology \& Gravitation,
  University of Portsmouth, Portsmouth, P01 2EG, U.K. \\ $^6$ South
  African Astronomical Observatory, P.O.\ Box 9, Observatory 7935,
  Cape Town, South Africa\\ $^7$ Department of Maths and Applied
  Maths, University of Cape Town, South Africa }

\maketitle 

\begin{abstract}
We optimize the design of future spectroscopic redshift surveys for
constraining the dark energy via precision measurements of the baryon
acoustic oscillations (BAO), with particular emphasis on the design of
the Wide-Field Multi-Object Spectrograph (WFMOS).  We develop a model
that predicts the number density of possible target galaxies as a
function of exposure time and redshift. We use this number counts
model together with fitting formulae for the accuracy of the BAO
measurements to determine the effectiveness of different surveys and
instrument designs. We search through the available survey parameter
space to find the optimal survey with respect to the dark energy
equation-of-state parameters according to the Dark Energy Task Force
Figure-of-Merit, including predictions of future measurements from the
Planck satellite.  We optimize the survey to test the $\Lambda$CDM
model, assuming that galaxies are pre-selected using photometric
redshifts to have a constant number density with redshift, and using a
non-linear cut-off for the matter power spectrum that evolves with
redshift. We find that line-emission galaxies are strongly preferred as targets 
over continuum emission galaxies. The optimal survey  covers a
redshift range $0.8 < z < 1.4$, over the widest possible area (6000
sq. degs from 1500 hours observing time). The most efficient number of 
fibres for the spectrograph is 2,000, and the survey performance 
continues to improve with the addition of extra fibres until a plateau is 
reached at 10,000 fibres.  The optimal point in the survey parameter space is not
highly peaked and is not significantly affected by including
constraints from upcoming supernovae surveys and other BAO
experiments.
 \end{abstract}

\section{Introduction}

The most exciting cosmological discovery in recent years has been the
observed late--time acceleration of the expansion of the Universe
(Riess et al.~1998; Perlmutter et al.~1999). When this observation is
combined with other cosmological data from the Cosmic Microwave
Background (CMB) and galaxy redshift surveys, the best fit model to
all these data is a flat Universe with $\simeq 30$\% of its energy
density in the form of dark matter and $\simeq 70$\% in the form of
``dark energy''; a mysterious component of the Universe with an
effective negative pressure.

The simplest explanation for dark energy is a ``cosmological
constant'' ($\Lambda$) as introduced by Einstein in his General Theory
of Relativity to create a static Universe. In recent times, the
cosmological constant has been interpreted as the ``vacuum energy'',
but the observed value of $\Lambda$ is a factor of $\sim10^{120}$
smaller than its natural value from particle physics (Krauss \& Turner
1995; Carroll 2001).  Alternatives to the cosmological constant
include time--evolving dark energy models, such as quintessence (Ratra
\& Peebles 1988; Wetterich 1988), and modifications of gravity at
large scales (Deffayet, Dvali \& Gabadadze 2002; Carloni et al. 2005; 
Damour, Kogan \& Papazoglou 2002).  The current cosmological
data is only good enough to measure the local value of the dark energy
density. Therefore, the present data prefers a cosmological constant,
mainly because it is the simplest model, whilst time--evolving models
of dark energy (which possess more free parameters) remain relatively
unconstrained by present datasets (see Corasaniti et al. 2004; Bassett
et al. 2004; Liddle et al. 2006).

Over the next decade, numerous experiments are proposed, over a wide
range of redshifts, to explore the time--evolution of dark energy and
determine its density as a function of cosmic time.  These experiments
(both ground--based and space--based) represent an investment of
billions of dollars and essentially use two general techniques to
probe the dark energy: geometrical tests of the expansion history of
the Universe using ``standard tracers'' (see below), and/or
observations of the rate of growth of large--scale structures
(clusters \& superclusters) in the Universe.  The relative merits of
these two techniques, and the proposed experiments, have been explored
in detail by the U.S. Dark Energy Task Force (DETF; Albrecht et
al. 2006) and a similar endeavour in the U.K. 

In this paper, we focus on measurements of the time--evolution of dark
energy using observations of the baryon acoustic oscillations (BAOs)
from large galaxy surveys. The BAOs are generated by acoustic waves in
the baryon--photon plasma in the early Universe, which become frozen
into the CMB radiation, and the distribution of matter, soon after the
Universe cools and re--combines at $z \simeq 1100$.  Over the last 5
years, the scale of these BAOs has been accurately measured in the CMB
by a number of experiments (WMAP, Archeops \& BOOMERanG), as well as
discovered in the distribution of matter (galaxies and clusters) at
low redshift (see Miller, Nichol \& Batuski 2001; Eisenstein et al.
2005; Cole et al. 2005; Padmanabhan et al. 2006). For example,
Eisenstein et al. (2005) used these observations to constrain the
flatness of the Universe to 1\%, under the assumption that dark energy
is a cosmological constant.

As the scale of the BAOs can be predicted to sub--percent accuracy
(see Eisenstein, Seo \& White 2006; Eisenstein \& White 2004), they
provide an excellent ``standard ruler'' which can be used to map the
geometry of the Universe through the angular-diameter distance and the
Hubble parameter relation (see Blake \& Glazebrook 2003; Seo \&
Eisenstein 2003; Hu \& Haiman 2003). Several spectroscopic experiments
have been proposed to observe and measure this standard ruler at high
redshift and thus constrain the time--evolution of dark energy, e.g.,
WFMOS (Bassett, Nichol \& Eisenstein 2005; Glazebrook et al. 2005a),
Baryon Oscillation Probe (Glazebrook et al. 2005b), VIRUS (Hill et
al. 2005) in the optical and NIR, and the Hubble Sphere Hydrogen
Survey (Peterson et al. 2006) and Square Kilometre Array (Blake et
al. 2004) in the radio. These new experiments share the desire to
measure millions of galaxy redshifts (at high redshift) over large
volumes of the Universe to control the errors from both cosmic
variance and Poisson noise (Blake \& Glazebrook 2003; Glazebrook \&
Blake 2005).

Given the large investment in time and money for these next generation
BAO experiments, we study here the optimal survey strategy for galaxy
redshift surveys like those proposed for WFMOS (Wide-Field
Multi-Object Spectrograph; see Glazebrook et al. 2005a).  This work
builds upon our previous development of the Integrated Parameter
Survey Optimization (IPSO) framework (Bassett 2005; Bassett, Parkinson
\& Nichol 2005) and addresses key observational issues such as:

\begin{itemize}
  
\item What are the optimal redshifts for BAO observations?
\item What is the best combination of areal coverage and target
  density?
\item What type of galaxies are the best targets?

\end{itemize}

Although our answers for these questions are derived for WFMOS--like
galaxy surveys, we believe that our results are general to BAO
experiments in optical wavebands.  The analysis presented in this
paper optimises $\Lambda$--cosmologies (i.e., a cosmological
constant) for the underlying cosmological model against which we are
optimizing.  Ideally we would like to optimize the observations for a
variety of dark energy models, and we will present such an analysis in
a future paper.  As a consequence therefore, our conclusions and
results naturally favour lower-redshift BAO observations where the
effect of $\Lambda$ on the expansion history of the Universe is
greatest.  We also neglect the improvements to the distance
measurements that will arise by reconstructing the acoustic peak in
the non-linear regime (see Eisenstein et al 2006b), assuming instead a
non-linear cut-off for the power spectrum that evolves with redshift.

In Section \ref{optimizingmethod}, we outline our methodology and
define the Figure-of-Merit (FoM) used to judge the optimality of
different survey designs.  In Section \ref{procedure}, we lay out the
procedures used to conducting the optimization, while in Section
\ref{LFs}, we describe the model we used to determine the density of
target galaxies. We present and discuss our results in Sections
\ref{results} and \ref{discussion} respectively. We conclude in
Section \ref{conclusions}.

\section{Optimizing Method}
\label{optimizingmethod}

We perform our optimization using the IPSO framework (Bassett 2005).
Consider a set of allowed survey geometries, indexed by $s$.  For each
survey geometry, $s$, we compute an appropriate Figure-of-Merit (FoM)
- also known as the utility in Bayesian evidence design, risk or
fitness - and optimization then simply proceeds by selecting the
survey geometry which extremises (minimising or maximising where
appropriate) the FoM.  Performing such an optimization therefore
requires three elements: a survey configuration parameter space $S$ to
search through, a target parameter space $\Theta$ of the parameters
that we wish to optimally constrain (labelled $\theta_{\mu,\nu...}$),
and a numerical Figure-of-Merit (FoM). We consider each of these three
elements in turn.

\subsection{Survey Parameters}

The survey parameters are those parameters which completely describe
the survey we wish to test. We start by splitting the survey into two
redshift regimes, one at low redshift ($z\simeq1$) and one at high
redshift ($z\simeq3$), and considering these regimes completely
separately. The properties of the survey in each regime are described
by a set of parameters: $\tau$ (survey time in that regime), $A$
(survey area in that regime), $n_{bin}$ (number of contiguous redshift
bins in that regime), $z_i$ (median redshift of the $i$th redshift
bin), and $dz_i$ (half-width of the $i$th redshift bin). There is also
a further parameter, $n_p$, which is the number of pointings performed
per field--of--view, i.e., the number of times the telescope observes
the same patch of the sky.  This parameter is derived from the other
parameters, using a model of the galaxy number density as a function
of exposure time, in conjunction with the assumed number of
fibres. All of these parameters are listed in Table
\ref{surveyparameters}.

\begin{table}
\centering
\caption{Survey parameters in each redshift regime}
\label{surveyparameters}
\begin{tabular}{cc} \hline
Survey Parameter & Symbol \\ \hline
Survey time & $\tau_{low}$, $\tau_{high}$ \\
Area covered  & $A_{low}$, $A_{high}$ \\
Number of redshift bins & $n_{bin}$ \\
Midpoint of  i$^{\rm th}$ redshift bin & $z_i$ \\
Half-width of i$^{\rm th}$ redshift bin & $dz_i$ \\
Number of pointings & $ n_p(low) $, $n_p(high)$ \\  \hline
\end{tabular}
\end{table}

In selecting a survey, we are also constrained by the technical
characteristics of the proposed instrumentation. In this paper, we
only consider a WFMOS--like experiment, whose important properties are
summarized in Table \ref{constraintparameters}. Firstly, there is the
total telescope time available, divided such that $\tau_{\rm Total} =
\tau_{high}+\tau_{low}$. There is also the field--of--view of the
telescope (FoV), which provides the amount of area that can be
observed per telescope pointing, and the number of fibres
($n_{fibres}$), which limits the number of objects that can be
observed simultaneously. We include the telescope aperture and fibre
diameter of the instrument, even though they are not direct constraint
parameters, because they affect the exposure time required to obtain
redshifts of galaxies. Note that, whilst this paper is mainly
concerned with finding the best survey for a given instrument, we can
also compare the performance of different instruments, which we do in
section \ref{sec:instruments}.

We include an ``overhead'' period between each observation, which is
considered as a period of ``dead'' time that is added on to the
exposure time to get the total time per pointing. We define a pointing
to be an exposure period to fill (or attempt to fill) all of the
fibres.  Even if the telescope is not actually moved between two
successive pointings on the sky, the fibres themselves would be
re--positioned (with the extra overhead time given in Table
\ref{constraintparameters}), and we would count this as two separate
pointings. We do not consider optimal tiling algorithms here.
Finally, we include limits on the minimum and maximum exposure time
per pointing.

We also impose constraints on the redshift regimes.  At low redshift,
there is no benefit in measuring the BAOs at $z<0.5$ because of the
recent measurements by Eisenstein et al.  (2005) and Cole et
al. (2005). There are also photometric-redshift survey observations of
the BAOs at $z\simeq0.5$ by Padmanabhan et al. (2006).  Therefore, we
constrain the ``low redshift'' regime in Table
\ref{constraintparameters} to fall within the range $0.5<z<1.5$.  The
upper limit derives from the wavelength coverage of the WFMOS
spectrographs, beyond which the [OII] emission line and 4000\AA\,
break shift out of the optical window, necessitating infra-red
spectographs.

We also consider a ``high redshift'' regime contained within the
limits $2.5<z<3.5$, motivated by observations of Lyman Break and
Lyman-$\alpha$ emission galaxies at $z\sim3$ (Steidel et al. 1999).
Spectroscopic redshifts for these galaxies are possible as the Lyman
Break and Lyman-$\alpha$ features are redshifted into the optical
window.

\begin{table}
\centering
\caption{Constraint parameters}
\label{constraintparameters}
\begin{tabular}{cc} \hline
Constraint Parameter & Value \\ \hline
Total observing time & 1500 hours \\
FoV & 1.5$^o$ diameter \\
$n_{fibres}$ & 3000 \\
Aperture & 8m \\
Fibre diameter & 1 arcsec \\
Overhead time & 10 mins \\
Minimum exposure time & 15 mins \\
Maximum exposure time & 10 hours \\
Low redshift range & $0.5 < z < 1.5$ \\
High redshift range & $  2.5 < z < 3.5 $ \\ \hline
\end{tabular}
\end{table}

\subsection{Target Parameters}

The target parameter space $\Theta$ is for our purposes the
cosmological parameter space.  The measurement of BAOs in the
tangential direction will allow us to measure the angular diameter
distance to a given redshift bin, whilst measurement of BAOs in the
radial direction will allow us to measure the Hubble expansion rate in
the redshift bin.  The angular-diameter distance (in a flat universe
containing only matter and dark energy) is given by,
\beq
d_{A}(z) = \frac{c}{H_0(1+z)} \int_0^z \frac{dz'}{E(z')} \,,
\eeq
where $E(z)$ is the normalised expansion as defined by,
\beq
E(z') = \sqrt{(\Omega_m (1+z')^3 + \Omega_{DE}f(z') )} \, ,
\eeq
where
\beq
f(z)=\exp{\int \frac{3(1+w(z))}{1+z}dz} \, ,
\eeq
and $w(z) = p/\rho$ is the equation of state parameter of the dark energy. In a flat
universe, $\Omega_{DE} = 1-\Omega_m$, which leaves only $H_0$,
$\Omega_m$ and $w(z)$ to be measured by any survey of the Universe.

For our optimization, we selected the Linder (Chevallier \&
Polarski, 2001, Linder, 2003) expansion of $w$ in
terms of the scale factor ($a$) of the Universe, rather than redshift,
as given by,
\beq
w(a) = w_0 + w_a(1-a) \,.
\eeq
We therefore optimize each survey for the values of $w_0$ and $w_a$,
whilst marginalizing over $H_0$ and $\Omega_m$.  For these latter two
parameters, we recast them as $\Omega_m h^2$ and $\Omega_m$ (where
$h = H_0/100$ km s$^{-1}$ Mpc$^{-1}$).

As discussed in the Introduction, we assume a cosmological constant
($\Lambda$) for the underlying cosmological model when constructing
our survey designs. However, we stress here that we still perform the
optimization using both the $w_0$ and $w_a$ parameters as outlined
above. This is the correct methodology as a priori we do not know the
true cosmological model that describes the Universe and must
therefore include the necessary parameters needed to describe the
range of models that possibly fit the observations. Furthermore, it
will allow us to compare these results to future predictions based on
a wider variety of dark energy models. We note the existence of
possible extensions to more complicated sets of target parameters,
such as those suggested by Albrecht \& Bernstein (2006).

\subsection{Figure of Merit}

The Figure-of-Merit (FoM) is a single real number assigned to each
survey configuration tested.  The survey with the best FoM is then
defined as the optimal survey.  The FoM we consider here is defined by
(Bassett 2004),
\beq
FoM(s_i) = \int_{\bf \Theta} I(s_i,\vec{\theta}) p(\vec{\theta})
d\vec{\theta}\, ,
\label{fom} 
\eeq 

\noindent where $I(s_i,\vec{\theta})$ is the performance of a survey
configuration ($s_i$), given a particular value of the target
parameters ($\vec{\theta}$) (here the dark energy parameters), and
$p(\vec{\theta})$ is a ``weighting vector'' that places emphasis on
particular volume of the parameter space.  By integrating over the
entire cosmological parameter space, the FoM we produce is only
dependent on the survey configuration.

Although there are many choices for $I(s_i,\theta)$ (Bassett 2005), we
follow Bassett, Parkinson and Nichol (2005) and choose D-optimality as
defined by 
\beq 
I(s,\theta) = \log \mbox{det}({\bf F} + {\bf P}), 
\label{dop} 
\eeq 

\noindent where ``det'' denotes the matrix determinant and ${\bf P}$
is the prior precision matrix (the Fisher matrix of all the relevant
prior data) and ${\bf F}$ is the Fisher information matrix of the
predicted likelihood, which we estimate using the method described by
Glazebrook \& Blake (2005).  From this, ${\bf F}+{\bf P}$ is the
Fisher matrix of the posterior and $\mbox{det}({\bf F} + {\bf P})$ is
inversely proportional to the square of the volume of the posterior
ellipsoid. We set ${\bf P}$ to zero, assuming no prior knowledge about
the dark energy. Therefore, a larger value for the FoM corresponds to
smaller errors on the parameters of interest. More details on how we
calculate the Fisher matrix are given in Appendix A.

In the US Dark Energy Task Force report (DETF; Albrecht et al. 2006b),
a slightly different Figure-of-Merit is used. Their FoM is defined to
be the reciprocal of the area in the $w_0$--$w_a$ plane that encloses
the 95\% C.L. region, whereas ours is the reciprocal of the square of
the area in the $w_0$--$w_a$ plane that encloses the 68\% region.
Therefore our FoM is proportional to the square of the DETF FoM.

\section{Optimization Procedure}
\label{procedure}

The computational problem of searching through the survey parameter
space ($S$) is that the available volume of this space is very large.
Therefore we use a Monte-Carlo Markov Chain (MCMC) procedure (the
Metropolis-Hastings algorithm; Metropolis et al. 1953; Hastings 1970)
to conduct a random walk through the parameter space to map the survey
parameters as a function of FoM, and find the optimal survey and the
survey configuration associated with it. As there may be a number of
degenerate minima in this parameter space (i.e., surveys with very
different configurations but similar FoMs), we also use the simulated
annealing algorithm (Kirkpatrick et al. 1983; Cerny 1985) to cool and
heat the chains generated by the MCMC process, such that we can be
sure we reach (or get very close to) the global maximum.

\begin{figure}
\center
\epsfig{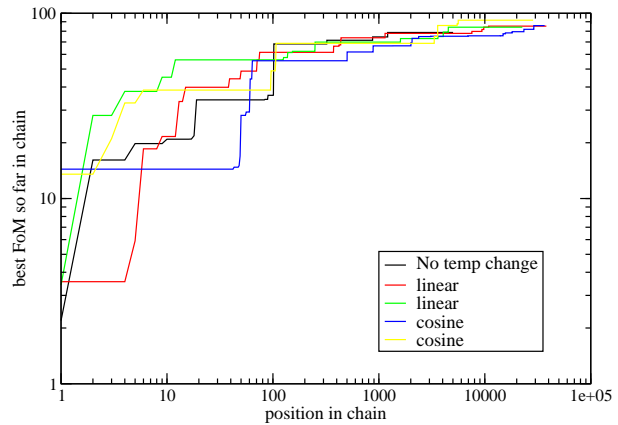}
\caption[schedule]{\label{schedule} The best Figure-of-Merit found by
  the MCMC process as a function of position in the chain for
  different thermodynamic schedules. These chains are taken from the
  optimization of surveys with two low redshift bins and one high
  redshift bin.}
\end{figure}

In MCMC, each new point in the chain is selected as a random point
near to the previous point with a probability equal to the ratio of
FoMs of the two corresponding surveys, such that a better survey will
usually be selected, whilst a worse survey will sometimes be
selected. In simulated annealing, the probability is multiplied by an
extra factor known as the ``temperature''.  When the temperature is
large, a wide variety of surveys are selected with almost equal
probabilities, but as the chains cool, and the temperature goes to
zero, the algorithm is weighted towards selecting only better surveys.
We cycle the temperature, using some thermodynamic scheduling, to
guarantee that we reach the global minimum in this large parameter
space.  In Figure \ref{schedule}, we show how these MCMC chain with
different thermodynamic schedules converge to the true optimal survey
configuration (the best FoM) after $\sim 10^4$ steps.

For this analysis, we judge the effectiveness of each survey
in terms of the cosmological constant ($\Lambda$) model and, as such,
the weighting function $p(\vec{\theta})$ from Eq.  \ref{fom} is set to
a delta-function around the best-fit $\Lambda$ model, 
\beq 
FoM({s_i})
= \int_{\bf \Theta} I(s_i,\vec{\theta}) \delta(\Lambda) d\vec{\theta}
= I(s_i,\Lambda) = \mbox{det}({\bf F}) \,.  
\label{finalfom}
\eeq

\noindent The optimization proceeds as follows:

\begin{enumerate}
\item Select a test survey configuration ($s$) from the given survey
  parameters (area coverage, redshift bins, exposure time etc.) from
  parameter space ($S$).
\item Estimate the number density of galaxies that will be observed by
  this test survey using luminosity functions described in Section
  \ref{LFs}.
\item Estimate the error on $d_A(z)$ and $H(z)$ using scaling
  relations given in Blake {\it et al.} (2006).
\item Calculate the Fisher matrix of parameters, using distance data
  plus other information that will be available from future
  experiments. This takes the form of Gaussian priors on $\Omega_m
  h^2$ and $\Omega_m$. For $\Omega_m h^2$ we use predicted Planck
  measurement with $\mu=0.147$ and $\sigma=0.003$. For $\Omega_m$ we
  assume $\mu = 0.3$ and $\sigma=0.01$. The procedure for this is
  outlined in Appendix A.
\item Use the Fisher matrix to calculate Figure-of-Merit for survey $s$,
  as given by Eq.\ref{finalfom}, using the $(w_0,w_a)$ submatrix.
\item Repeat steps 1 through 5, conducting a MCMC search over the
  survey configuration parameter space $S$, attempting to minimize the
  FoM.
\end{enumerate}

There are a number of assumptions that we make in doing this
optimization that may bias the results in some manner. They are: (1)
we are applying a constant number density with redshift using
photometric-redshift target pre-selection in order to ensure the most
efficient use of fibres, (2) we are optimizing for $\Lambda$CDM, and
(3) we are neglecting reconstruction of the acoustic peak at
non-linear scales (Eisenstein et al 2006b) (i.e. assuming a non-linear
power spectrum cut-off which evolves with redshift in the same style
as Glazebrook \& Blake 2005).

The assumed fiducial cosmological  parameters are $\Omega_m =0.3$,
$H_0 = 70{\rm kms^{-1}Mpc^{-1}}$, $\Omega_mh^2 = 0.147$, and of course 
$w_0=-1$ and $w_a=0$. We also assume a baryon fraction of 0.15, giving 
$\Omega_b = 0.045$ and a sound horizon of $s = 108h^{-1}{\rm Mpc}$.

When optimizing a system, it is important to know how sensitive the
optimal solution is to a slight change in the parameters.  We quantify
this sensitivity using ``flexibility bounds'', which are defined as
the change in a survey parameter which decreases the FoM by 20\%
relative to the optimum. The value of 20\% is arbitrary, based on
comparisons of survey performance with other planned experiments and
possibilities of theoretical advancements.  What this will mean for
the error bars on the two parameters ($w_0$, $w_a$) is that, since the
FoM is the square of the 1-$\sigma$ error ellipse, a 20\% decrease in
the FoM corresponds to a 10\% increase in the size of the error
ellipse, or a $\approx 5\%$ increase in each error bar.  We computed
the flexibility bounds by considering the effects of changing only one
parameter, whilst keeping the other parameters at their optimal values
(we do not account for correlations between parameters in their FoM
dependence).

\section{Galaxy number counts modelling}
\label{LFs}

We describe here our model for converting a survey exposure time and
redshift range into an observed number density of target galaxies for
spectroscopy. The number counts calculator is constructed using the
observed properties of $z \sim 1$ and $z \sim 3$ galaxies taken from
the literature (e.g. observed luminosity functions and distributions
of equivalent widths of emission lines).  We do not attempt to model
observational issues such as the reliability of redshift extraction or
confusion in line identifications.

We consider four different classes of galaxy:

\begin{itemize}
  
\item The pre-selection of $z \sim 1$ ``red'' (passive) galaxies using
  multi--colour photometric data. Spectroscopic redshifts for these
  galaxies would be measured using strong continuum features such as
  the 4000\AA\, break. This is similar in spirit to the Luminous Red
  Galaxy (LRG) selection of Eisenstein et al. (2001).
  
\item The pre-selection of $z \sim 1$ ``blue'' (star-forming) galaxies
  using multi--wavelength imaging data (e.g. GALEX+SDSS).
  Spectroscopic redshifts for these galaxies would be measured from
  the [OII] emission-line doublet, which is expected to be strong at
  $z \sim 1$ owing to the evolution of the cosmic star formation rate
  (Willmer et al.\ 2005; Cooper et al.\ 2006).
  
\item Spectroscopic follow-up of Lyman Break Galaxies (LBGs) at $z
  \sim 3$, obtaining the redshift from continuum features.
  
\item Spectroscopic follow-up of LBGs at $z \sim 3$, but obtaining the
  redshifts from the Lyman-$\alpha$ emission line where possible.
  This allows for much shorter exposures, but results in a lower
  fraction of galaxies with spectroscopic redshifts as only a fraction
  of LBGs are expected to have strong emission lines.  The remaining
  fibres are assumed to produce failed redshifts and are thus not
  included in the BAO analysis.

\end{itemize}

In Table \ref{tabsn}, we list our assumptions for the required
signal-to-noise ratio ($S/N$) to obtain a successful redshift for the
four different galaxy classes discussed above. These estimates are
based on previous experience from the literature and the 2dF--SDSS LRG
and QSO (2SLAQ) survey (Cannon et al. 2006).

\begin{table}
\begin{center}
\caption{Signal-to-noise requirements for a successful redshift,
  defined at the redshifted [OII] or Lyman-$\alpha$ wavelength for
  emission-line galaxies, and in the $R$-band for red galaxies.  The
  line is assumed to be unresolved in a 1 nm spectral resolution
  element.}
\begin{tabular}{lll}
\hline
Galaxy class & $S/N$  \\
\hline
$z=1$ ``red'' (passive) galaxies & 7 per nm  \\
$z=1$ ``blue'' (emission-line) galaxies  &  7  \\
$z=3$ LBGs (continuum redshift) &  7 per nm \\
$z=3$ LBGs (Ly-$\alpha$ redshift) &  5  \\
\hline
\label{tabsn}
\end{tabular}
\end{center}
\end{table}

For a given exposure time and redshift, the number counts calculator
determines the flux limit reachable for emission lines ([OII] or
Lyman-$\alpha$), or the apparent magnitude in the $R$-band reachable
for continuum redshifts, using a full photon-counting calculation
assuming the $S/N$ values listed in Table \ref{tabsn}.  The WFMOS
instrument is assumed to have 1-arcsec diameter fibres (with a factor
$0.7$ ``aperture light loss'') and to be mounted on a telescope with
an 8-m diameter collecting area.  In addition we assume an overall
spectrograph efficiency of $10\%$, and that the emission line is
unresolved in a 1 nm spectral resolution element.  The nod-and-shuffle
mode of operation is assumed, which implies that a factor of 4
increase in integration time is necessary to achieve the same
signal-to-noise ratio in the spectrum (for background-limited
observations), with much-reduced sky-subtraction systematics.  The
brightness of the sky (in AB magnitudes per arcsec$^2$) is assumed to
increase with wavelength $\lambda$ as:

\beq
{\rm Sky}_{AB}(\lambda) = 22.8 - \frac{(\lambda{\rm(\AA)}-4000)}{1500}
\eeq

\noindent We do not include the very rapid variation in sky brightness
with wavelength caused by individual sky lines, as this would add
greater complexity to the optimization, and furthermore each candidate
survey involves a much broader redshift range.

The number counts calculator demonstrates that WFMOS can detect a
$z=1$ line with flux $3 \times 10^{-20}$ W m$^{-2}$ at $S/N = 10$ in
30 mins integration, and a $z=1$ continuum galaxy with $R_{AB} = 23.6$
with $S/N = 6$ per nm in 40 mins integration.  More details are
available in the WFMOS Feasibility Study (Barden et al. 2005); these
numbers are consistent with existing high--redshift spectroscopic
surveys on 8-metre class telescopes (DEEP, VVDS).

Having established the galaxy flux limit in line or continuum
emission, we then model the galaxy luminosity function as a Schechter
function.  For the $z \sim 1$ galaxy samples we use the results of the
DEEP redshift survey (Willmer et al.\ 2005) separately for the ``red''
and ``blue'' galaxies.  For ``blue'' galaxies, we scale $L^*$ with
redshift using $L^* \propto (1+z)^3$, to model the evolution of the
cosmic star--formation rate (see Hopkins 2004).  For the $z \sim 3$
samples, we use the results from photometric-redshift surveys in the
Hubble Deep Fields described by Arnouts et al.\ (2005).  These latter
results do not differ significantly from those reported by Steidel at
al.\ (1999). The parameters of the galaxy luminosity functions are
summarized in Table \ref{tablum} and assume $\Omega_{\rm m} = 0.3$,
$\Omega_\Lambda = 0.7$ and $h = 0.7$.  For the DEEP survey luminosity
function, we use the ``minimal weights'' to be conservative.

\begin{table*}
\begin{center}
\caption{Galaxy luminosity function parameters.  The Schechter
  function is defined as $dn/dL = \phi^* \, (L/L^*)^\alpha
  \exp{(-L/L^*)}$ where, in terms of the absolute magnitude $M$,
  $L/L^* = 10^{0.4(M^* -M)}$.  For $z=1$ blue galaxies, $L^*$ is
  assumed to scale with redshift as $(1+z)^\beta$ and its value is
  quoted at $z = z_0$.}
\begin{tabular}{llcccccl}
\hline

Type & Magnitude & $M^*(z=z_0)$ & $\phi^*$ (Mpc$^{-3}$) & $\alpha$ &
$z_0$ & $\beta$ & Reference \\

\hline

$z=1$ blue & $B_{\rm Vega}$ & $-21.38$ & $2.08 \times 10^{-3}$ &
$-1.3$ & $1.1$ & $3.0$ & Willmer et al.\ (2005) Table 4 \\

$z=1$ red & $B_{\rm Vega}$ & $-21.11$ & $1.07 \times 10^{-3}$ & $-0.5$
& --- & --- & Willmer et al.\ (2005) Table 5 \\

$z=3$ & $AB (\lambda = 6060$\AA) & $-21.08$ & $1.62 \times 10^{-3}$ &
$-1.47$ & --- & --- & Arnouts et al.\ (2005) Table 1 \\

\hline
\label{tablum}
\end{tabular}
\end{center}
\end{table*}

For the emission-line samples, we assume that galaxies in the given
redshift range are targetted by fibres down to a faintest continuum
magnitude consistent with the fibre density.  In order to estimate the
fraction of targetted galaxies surpassing the limiting line flux, we
combined the continuum flux distribution (from the luminosity
functions) with the observed equivalent width distributions of [OII]
(for the $z \sim 1$ ``blue'' sample) and Ly$\alpha$ (for the $z \sim
3$ LBGs).  We neglect any correlations between the equivalent width of
these lines and the galaxy luminosity, and simply apply the same
equivalent width distribution as we integrate over continuum flux.  We
also make no attempt to optimize our selection by concentrating, for
example, on the bluest galaxies (apart from our initial division of
the galaxy population into ``blue'' and ``red'' populations). For
[OII], we estimate the equivalent width distribution from the DEEP
redshift survey, whilst for Ly$\alpha$ we use the work of Shapley et
al.\ (2003).  Our assumptions are given in Table \ref{tabline}, where
the equivalent width distributions are given in the observed frame.
In Table \ref{tabmisc}, we list for completeness the magnitude
conversions and K-corrections used in our analysis.

\begin{table*}
\begin{center}
\caption{Equivalent width distributions for emission lines}
\begin{tabular}{lll}
\hline
Line & Distribution & Reference \\
\hline

${\rm [OII]}$ at $z=1$ & Gaussian with mean 80\AA\, and
r.m.s.\ 40\AA\, & DEEP survey (priv. comm.) \\

Ly-$\alpha$ at $z=3$ & Fraction $>W = 0.14 \times \ln{(100/W)}$ where
$W > 5$\AA & Shapley et al.\ (2003) \\ \hline

\label{tabline}
\end{tabular}
\end{center}
\end{table*}

\begin{table*}
\begin{center}
\caption{Magnitude conversions and K-corrections}
\begin{tabular}{lll}
\hline
Type & Conversion & Reference \\
\hline

Filter systems & $R_{\rm Vega} = R_{AB} - 0.218$ & Willmer et
al.\ (2005) Table 1 \\

& $B_{\rm Vega} = B_{AB} + 0.098$ & Willmer et al.\ (2005) Table 1 \\

Colour ($z=1$, blue) & $U_{AB} - B_{AB} = 0.7$ & DEEP survey
(priv. comm.) \\

K-correction ($z=1$, blue) & $B_{\rm rest} - R_{\rm obs} = (z - 0.5) -
0.9$ & Willmer et al.\ (2005) Figure A15 \\

K-correction ($z=1$, red) & $B_{\rm rest} - R_{\rm obs} = 2 (z - 0.5)
- 0.9$ & Willmer et al.\ (2005) Figure A15 \\

K-correction ($z=3$) & Use LBG template spectrum & \\

\hline
\label{tabmisc}
\end{tabular}
\end{center}
\end{table*}

\begin{figure}
\begin{center}
\epsfig{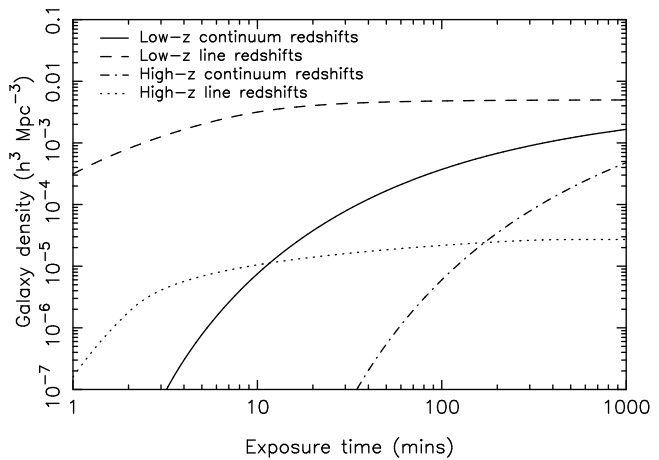}
\caption{Galaxy number densities obtained for the four galaxy classes
  discussed in the text as a function of exposure time (in a single
  telescope pointing), assuming 5000 fibres.  The curves all reach a
  plateau when the surface density of galaxies reach the spectrograph
  fibre density as given in Table 1.}
\label{numbercountsfig}
\end{center}
\end{figure}

In Figure \ref{numbercountsfig} we plot the average number density of
galaxy targets for the four galaxy classes discussed above, as a
function of exposure time.  For the two $z \sim 1$ samples, we assume
a target redshift range of $0.9 < z < 1.1$.  For the two $z \sim 3$
samples, we assume a target redshift range of $2.9 < z < 3.1$.  We
note that in order to achieve ``cosmic-variance limited'' surveys
(i.e., the error on the cosmological parameters is not dominated by
Poisson noise in the power spectrum), we require a number density in
the range $10^{-4} \rightarrow 10^{-3} \, h^3$ Mpc$^{-3}$, depending
on the clustering properties of the galaxies.  Observations of $z \sim
1$ line-emitting (``blue'') galaxies reach this requirement in very
short exposure times of a few minutes for our spectrograph and $S/N$
assumptions.  The disadvantage of these targets is their relatively
low galaxy bias relative to the ``red'' sample, and that photometric
redshift pre-selection is much harder to achieve because of the large
dispersion in their intrinsic colours (compared to passively--evolved
ellipticals for example).  $z \sim 1$ continuum redshifts require at
least 30 minutes integration.  $z \sim 3$ continuum redshifts demand
exposure times of about 4 hours, and surveys using $z \sim 3$ line
redshifts will have a significant Poisson noise component.

For a broad $z \sim 1$ redshift bin, we apply the number counts
calculator to the deepest redshift slice of width $\Delta z = 0.2$,
which requires the longest integration time.  We then assume that
photometric pre-selection is used to create a constant number density
of targets throughout the broader redshift bin, such that galaxies at
lower redshifts in the bin will typically have higher $S/N$ than the
minimum values listed in Table \ref{tabsn}.

\section{Survey optimisation}
\label{results}

We now apply the procedure outlined in Section \ref{procedure} to find
the optimal survey configuration. For each case specified, we run the
MCMC search through the parameter space using several simulated
annealing cycles of heating and cooling, with a single heating and
cooling cycle defined as 5000 MCMC steps. We run several parallel
chains to judge the convergence of the chains on the optimal solution.

\subsection{Optimal galaxy population}
\label{galaxypop}

We first assess the optimal galaxy population to target for BAO
measurements using a WFMOS--like instrument. This is achieved by
studying the trade-off between exposure time and areal coverage for
the four galaxy classes considered.  For the $z=1$ ``blue''
(emission-line) galaxies, we assume a galaxy bias factor (relative to
the underlying dark matter power spectrum) of $b=1.3$, whilst for the
$z=1$ ``red'' (passive) galaxies, we assume $b=2$.  We assume $b=3$
for both of the $z=3$ galaxy samples considered.  We do not attempt to
model the dependence of bias on redshift or luminosity.

We consider surveys where galaxies are observed at both high and low
redshifts, and the FoM comes from the combination of the two redshift
bins.  For this initial study we fix the two redshift regimes at $0.5
< z < 1.3$ for ``low redshift'' galaxies ($z_{low} =0.9$ and
$dz_{low}=0.4$) and $2.5 < z < 3.5$ for ``high redshift'' galaxies
($z_{high}=3.0$ and $dz_{high}=0.5$).  We also fix the total time
spent observing in each regime, $\tau_{low}=$ 800 hrs and
$\tau_{high}=$ 700 hrs.  We vary only the areas and exposure times for
surveys based on these four galaxy classes.  Note that if the total
survey time is fixed, then the exposure times and areas are linked
through the number of repeat pointings for each field-of-view. This
parameter ($n_p$) is selected by an algorithm that makes best use of
the available number of fibres.

\begin{figure}
\center
\epsfig{file=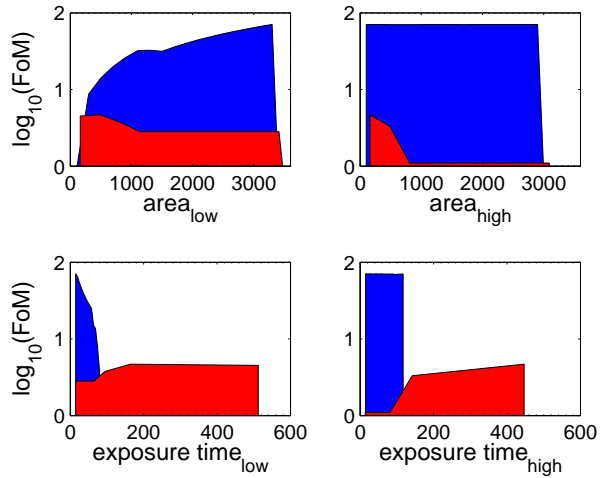, width=8cm}
\caption[areatime]{\label{areatime} The distribution of survey
  parameters by FoM for the area (in sq.\ degrees, top) and exposure
  time (in minutes, bottom) for surveys observing 800 hours in the low
  redshift bin in combination with 700 hours in the high redshift
  bin. The shaded region delimits allowed surveys. The surveys observe
  either line (blue) or continuum (red) galaxies. The area of the low
  redshift bin is well determined (3393 sq.\ degs for line, 371 for
  continuum), but the area of the high redshift bin is not determined
  at all because this bin does not add to the FoM.}
\end{figure}

The results of this study are presented in Table \ref{galaxytypetable}
where we list the optimal survey for each of the four possible
combinations of the four galaxy classes considered, i.e., (1)
emission-line galaxies in both the low and high redshift regimes, (2)
continuum (or passive) galaxies at low redshift and emission-line
galaxies at high redshift, (3) emission-line galaxies at low redshift
and continuum galaxies at high redshift and (4) continuum galaxies at
both redshifts.  We provide the optimal areas and exposure times in
the table (with their flexibility bounds) as well as the overall FoM
for the optimal survey. The Figure-of-Merit as a function of galaxy
type, and area and exposure times for the low and high redshift bins,
are plotted in Figure \ref{areatime}. The Figure-of-Merit for the high
redshift surveys implicitly includes the best low-z component, and
vice versa.

\begin{table*}
\centering
\caption{Optimal survey parameters with their flexibility bounds in
  brackets for the four galaxy classes discussed in the text. Surveys
  spend 800 hours at low redshift and 700 hours at high redshift.}
\label{galaxytypetable}
\begin{tabular}{lcccc} \hline
  &  \multicolumn{4}{c}{Survey}  \\
   Survey Parameters & 1 & 2 & 3 & 4  \\ \hline
   Low-z bin & Line & Cont & Line & Cont\\
-  $A_{\rm low}$ (sq. degs) & 3393 (2700-3393) & 750 (0-750)  & 3382 (2700-3393) & 438  (100-2300)  \\
    - exp. time (mins) & 15 (15-21) &  204 ($>60$) & 15 (15-27) & 180  ($>25$) \\
  - num. dens $\times 10^4$ ($h^3$ Mpc$^{-3}$) & 8.0  & 1.5 & 8.0 & 0.4  \\
  High-z bin & Line & Line & Cont & Cont \\
  -  $ A_{\rm high}$ (sq. degs) & 2900 (0-2968) & 2945 (1700-2967) & 140 (0-2000) & 123 (100-2000) \\ 
  - exp time (mins) & 15 ($>15$) &  15 (15-90) & 530 ($>27$) & 25 ($>25$) \\
  - num. dens $\times 10^4$ ($h^3$ Mpc$^{-3}$) & 0.2 & 0.2 & $2.2$  & $ 2.5$  \\ 
   FoM & 70.6 &  3 & 70.5 & 3  \\ \hline
 
\end{tabular}
\end{table*}

Table \ref{galaxytypetable} demonstrates that the $z=1$ ``blue''
(emission-line) galaxy population does significantly better than the
$z=1$ ``red'' (passive) population, as the FoM for these ``blue''
galaxies (in case 1) is a factor of 25 higher (than in case 2). This
corresponds to a factor of $\sqrt{25} \simeq 5$ improvement in the
area of the error ellipse in the $w_0$--$w_a$ parameter space. For
circular error ellipses, this is a factor of $\simeq 2.2$ improvement
in each parameter. This advantage is due to the speed at which
redshifts can be obtained for these $z=1$ ``blue'' galaxies (from the
[OII] emission line), which allows the survey to cover more area per
unit time compared to targetting ``red'' galaxies (even though these
red galaxies have a higher bias factor).

Table \ref{galaxytypetable} demonstrates that observing either line or
continuum galaxies at high redshift does not improve the FoM, no
matter what type of galaxies are observed at low redshift.  This
manifests itself in the flatness of the FoM plots for the high
redshift bin area and exposure time in Figure \ref{areatime}, and the
corresponding width of the flexibility bars.  The ineffectiveness of
the high-redshift bin is due to the small energy density of the dark
energy at this redshift for our assumed cosmological constant fiducial
model, and may change for more general dark energy models.

\subsection{Optimal redshift regime}

In the previous subsection, we held fixed the limits of the two
redshift regimes and the total time spent in each, whilst varying the
area and exposure times of surveys for the four different galaxy
samples.  Here, we address the issue of the optimal redshift range for
constraining dark energy models by allowing the total times
($\tau_{low}$ and $\tau_{high}$), the central redshift ($z_i$), and
width ($dz_i$), of the two redshift regimes to vary as free parameters
in our MCMC search. We still however impose the global redshift
constraints of $0.5<z<1.5$ and $2.5<z<3.5$ to ensure we obtain
realistic survey configurations.

\begin{figure}
\center
\epsfig{file=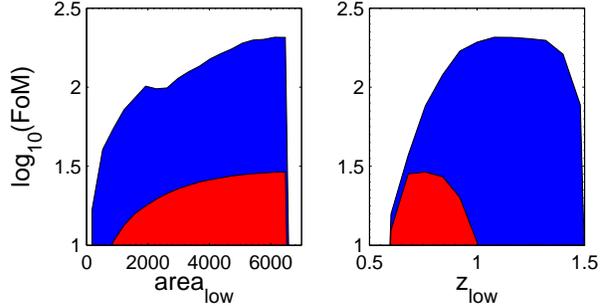, width=8cm}
\caption[singlelowzbin]{\label{singlelowzbin} The distribution of
  survey parameters by FoM for the area and midpoint redshift for a
  single bin at low redshift for line-emission (blue) and continuum
  (red) strategies . The shaded region delimits allowed surveys. Note
  that since there is only one bin, all time is spent observing at
  this redshift (1500 hours).}
\end{figure}

\begin{figure}
\center
\epsfig{file=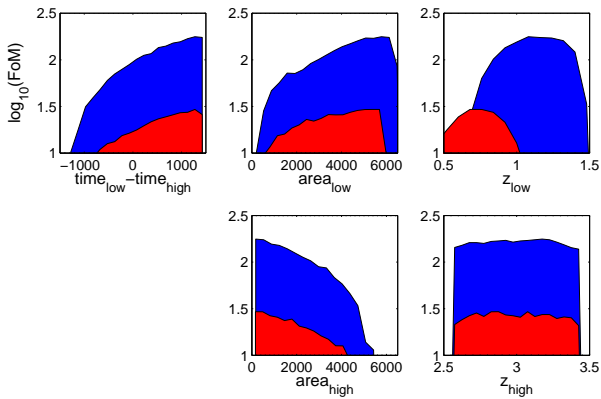, width=8cm}
\caption[twobins]{\label{twobins} The distribution of survey
  parameters by FoM as for Fig. \ref{areatime}, with the position and
  widths of the two redshift bins being allowed to vary. Note that the
  red colour indicates the galaxy type at low redshift only
  (continuum), whereas the high redshift bin targets line-emission
  galaxies in both cases.}
\end{figure}

The results of this search are presented in Figures
\ref{singlelowzbin} and \ref{twobins}, as well as in Table
\ref{binredshifttable}. First, we consider a single ``low redshift''
bin (surveys A for ``blue'' galaxies and C for ``red'' galaxies in
Table \ref{binredshifttable}), where the mean redshift and redshift
widths are allowed to move.  Note that now all the survey time is
spent at low redshift ($\tau_{low} = 1500$ hrs). As seen in Figure
\ref{singlelowzbin}, these changes in survey time and redshift limits
have a marked effect on the FoM, i.e, 218 compared to 72 for blue
galaxies and 42 compared to 1.8 for red galaxies. The area of the
survey increases in both cases, whilst the optimal redshift increases
for the blue galaxies ($z_{low} = 0.9 \rightarrow 1.1$), and decreases
for the red galaxies ($z_{low} = 0.9 \rightarrow 0.8$).

We next consider the question of whether a high redshift bin adds to
the Figure-of-Merit, and what the optimal splitting of total survey
time between the two bins is. We consider the case of two redshift
bins (``high'' and ``low''), where the high redshift galaxies are
observed by line emission, and the total survey time is a constraint
(i.e. $\tau_{low} + \tau_{high} = \tau_{total} = 1500$ hrs). The
results are presented in Table \ref{binredshifttable} and Figure
\ref{twobins}.

The addition of a ``high redshift'' bin for line-emission galaxies
(surveys B and D in the table) does not improve the FoM, it in fact
gets slightly worse, 176 compared to 207 for a single low-redshift bin
of line emission, or doesn't change, 29 compared to 29 for a single low-redshift bin
of continuum emission.

\begin{table*}
\centering
\caption{Best fit survey parameters allowing the redshift bins to move
  (without and with a line-emission redshift bin at high redshift),
  with the flexibility bounds in brackets}
\label{binredshifttable}
\begin{tabular}{lcccc} \hline
  &  \multicolumn{4}{c}{Survey}  \\
   Survey Parameters & A & B & C & D  \\ \hline
   Low-z bin & Line & Line & Cont & Cont \\
   -  $A_{\rm low}$ (sq. degs) & 6300 (4500-6300) & 5600 (4500-6150)  &  6300 (3700-6300) & 5448 (3100-6000)\\
  - $\tau_{\rm low}$ (hrs) & 1500 (-) & 1400 (1100-1500) & 1500 (-) &  1400 (1100-1500)\\
  - $z_{\rm low}$ & 1.1 (0.95-1.35) & 1.08 (0.95-1.35) & 0.74 (0.65-0.9) & 0.72 (0.65-0.9)\\ 
  - $dz_{\rm low}$ & 0.3 (0.12-0.5)  & 0.35 (0.14-0.48)   & 0.18 (0.11-0.26)   &  0.19   (0.11-0.26)   \\
    - exp. time (mins)  & 15 & 19 & 15 & 17 \\
  - num. dens $\times 10^4$ ($h^3$ Mpc$^{-3}$) & 6.5 & 7.1 & 3.7 & 4.5  \\
  High-z bin & - & Line & - & Line \\
  -  $ A_{\rm high}$ (sq. degs) & - & 150 (0-1200) & - & 43 (0-1200) \\
  - $\tau_{\rm high}$ (hrs) & - & 100 (0-400) & - & 100 (0-400) \\
  - $z_{\rm high}$& - &  3.15 (2.6-3.4) & - & 2.9  (2.65-3.35) \\ 
  - $dz_{\rm high}$ & -   & 0.13 (0.1-0.5)   & -   & 0.27 (0.1-0.5)  \\
  - exp time (mins) & - & 60 &-  & 240 \\
  - num. dens $\times 10^4$ ($h^3$ Mpc$^{-3}$) & - & 0.13 & - & 0.35 \\
   FoM & 207 & 176  &29 & 29 \\ \hline
\end{tabular}
\end{table*}

\subsection{Optimal number of redshift bins}

The analysis presented in Section 5.2 demonstrated the sensitivity of
the FoM to both the location, and number, of redshift regimes, i.e.,
our optimization clearly preferred a single ``low redshift'' bin.
Therefore, we study here the benefits of splitting this single ``low
redshift'' bin into multiple thinner redshift bins, assuming these
thinner bins are contiguous over the full redshift range of the single
wider bin. We expect these thinner redshift bins to have worse
measurements of $d_A(z)$ and $H(z)$, compared to the single wider bin
(because their volumes will be smaller), but this disadvantage could
be overcome by the increased number of distance measurements available
for fitting the cosmological models (we do not consider correlations
between the errors of different redshift bins). Therefore, we
performed a MCMC search as a function of the number of ``low
redshift'' bins and discovered that the FoM quickly saturated at
$\simeq260$ for all configurations with more than one ``low redshift''
bins, i.e., the optimal survey would split the single low-redshift
regime into two equally--sized bins.  These results should be
revisited for different dark energy models.

\subsection{Global Optimum}

We find that the best Figure-of-Merit searching through all possible
survey configurations is one which spends all its time at low redshift
surveying an area of around 6000 sq. degs, targetting line-emission
galaxies (survey A in Table \ref{binredshifttable}).  The medium
redshift of the bin is about 1.1 and it will stretch from $z \sim 0.8
- 1.4$.  The exposure time is 15 minutes per field-of-view.  A
additional high redshift bin is disfavoured.

This optimum is not highly peaked, as we can see by looking at Figure
\ref{singlelowzbin}. The flattening of the FoM curve for area at about
6000 sq.\ degs and the flat plateau at the top of the $z_{low}$ curve
indicates that deviations from these optimal values will result in
only small changes in the Figure-of-Merit, also shown by the large
width of the flexibility bars in Table \ref{binredshifttable}. This
shows that these results are robust against moderate changes in the
survey design.

\section{External Constraints}
\label{discussion}

In the previous Section we focussed on optimizing the survey
parameters to obtain the best constraints on the properties of dark
energy, whilst keeping the constraint parameters fixed. We now
consider the case where the current constraint parameters are changed
or new constraints are added, and the resulting effects on the optimal
survey FoM and configuration.

The constraint parameters cannot be considered as simple survey design
parameters, as they are built in at an instrument level.  Furthermore,
the constraint parameters will be unbounded by FoM in one
direction. For example a survey that runs for 5 years will always do
better than a survey that lasts for only 3 years. However, the
behaviour of the FoM will be different for different constraint
parameters. The Figure-of-Merit of the best survey will continue to
scale as the total survey time is increased, but may quickly asymptote
to some maximal value in the case of the number of spectrograph
fibres.

We consider three cases: (1) a extra constraint is imposed on the
maximum area to be surveyed, motivated by the size of the input
photometric catalogue; (2) the number of spectroscopic fibres is
changed; and (3) the telescope aperture and field-of-view are changed.
Finally, we consider how the optimal design changes if constraints
from other dark energy surveys (of both supernovae and baryon acoustic
oscillations), which will have been completed by the time that a
WFMOS-like instrument is constructed, are included in the analysis.

\subsection{Input Photometric Catalogue}

The optimal surveys are those with the maximum possible area,
surveying thousands of square degrees.  However, a spectroscopic
survey requires an input catalogue of photometrically pre-selected
galaxies.  How would the dark energy constraints be affected if the
available area of the input catalogue is less than the optimal value?

Looking at Table \ref{binredshifttable} for a single redshift bin, we
see that the flexibility bounds place a lower limit on the total
survey area of 3000 sq. degs., compared to the optimal area of 6000
sq. degs.  Such imaging surveys, whilst not yet in existence, are in
the planning stages (for example the Dark Energy Survey (DES), see
Abbott {\it et al.} 2005).

\subsection{Number of fibres}
\label{sec:fibres}

\begin{figure}
\center
\epsfig{file=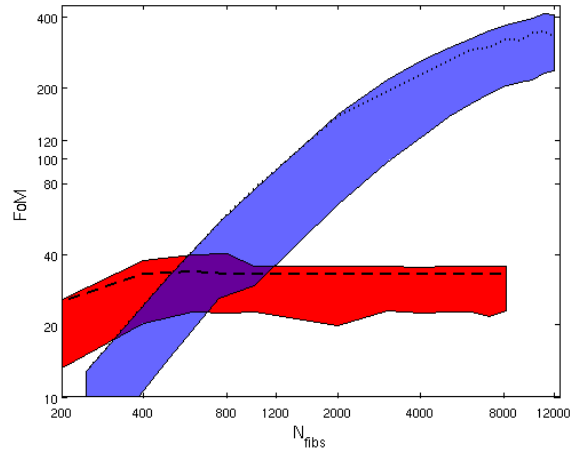, width=8.5cm}

\caption[fibres]{\label{fibres} The variation of the FoM of the
  optimal survey with the total number of spectroscopic fibres, for
  both line-emission galaxies (dotted curve) and continuum galaxies
  (dashed curve), for a single bin at low redshift. The line-emission
  curve does not reach an asymptopic limit before 10{,}000 fibres,
  whereas the continuum curve reaches its asymptotic value at around
  750 fibres.  We also include the effect of varying the number counts
  models of the galaxies in a band between ``pessimistic'' and
  ``optimistic'' limits (line-emission in blue and continuum-emission
  in red).  This extra freedom does not change the results concerning
  number of fibres.  For a small number of fibres, the optimistic and
  default models give the same FoM, as the optimal surveys have
  reached the maximum number of pointings and smallest exposure time,
  and no more galaxies can be observed.  We assume that fibres can
  access any part of the field-of-view.}
\end{figure}

In this Section we investigate the variation of the optimal
Figure-of-Merit with the available number of spectrograph fibres,
assuming a single survey bin at low redshift.  We consider both line
and continuum-emission survey strategies, including both
``pessimistic'' and ``optimistic'' versions of the number counts
calculation, which correspond simply to decreasing and increasing the
predicted number of $z \sim 1$ galaxies by 50\%.  We assume that the
fibres are able to access any part of the field-of-view. We note that
this is more challenging for Echidna systems (Kimura {\it et al.}
2003), which have a fairly restricted ``patrol radius'' for each
fibre. The results can be seen in Figure \ref{fibres}.

The optimal value of the number of fibres for line-emission strategies
is around 10{,}000 for a single low redshift bin.  At this point the
number density of galaxies sampled is more than sufficient that shot
noise is negligible. The optimal value for the continuum case is about
750.  We note that the large difference in fibre numbers between the
two cases is due to the ability of the line-emission survey to access
higher redshifts.  These optimal numbers of fibres are not affected by
shifting to a ``pessimistic'' or ``optimistic'' number counts model.

\begin{figure}
\center
\epsfig{file=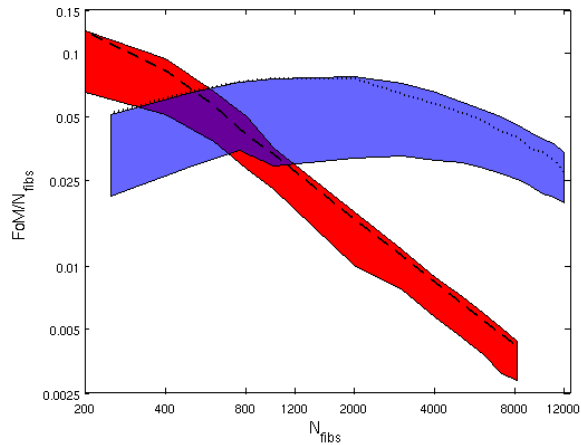, width=8cm}
\caption[fibresefficiency]{\label{fibresefficiency} Constructed from
  the same data as Figure \ref{fibres}, but we now plot
  FoM/$n_{fibres}$ against the number of fibres, to find the most
  efficient use of the instrument.}
\end{figure}

We note that the optimal survey configuration for 10{,}000 fibres only
obtains successful redshifts for about 6000 galaxies, with the
remaining targets possessing emission-line fluxes that fall below the
minimum $S/N$ level and produce failed redshifts.  Therefore, it is
also interesting to consider the most efficient use of fibres.  One
method of quantifying this ``fibre efficiency'' is to plot the FoM per
fibre as a function of fibre number, as shown in Figure
\ref{fibresefficiency}.  For line-emission galaxies, the most
efficient number of fibres judged by this statistic is 2000, but this
survey configuration has a significantly lower FoM compared to the
optimal survey (151, compared to 328 for 10{,}000 fibres). A happy
medium would be a total of 3000-4000 fibres.

\subsection{Telescope Aperture and Field of View}
\label{sec:instruments}

In this Section we consider executing the galaxy surveys using a
spectrograph differing from the WFMOS specifications presented in
Table \ref{constraintparameters}.  We consider three alternate
possibilities: reducing the WFMOS field-of-view from $1.5$ deg to
$1.0$ deg diameter, using the AAOmega spectrograph on the
Anglo-Australian Telescope (AAT), and using the Sloan Digital Sky
Survey (SDSS) hardware.

The reduction of the field-of-view is a simple alteration, as it does
not affect the exposure time required to obtain the redshift of a
galaxy (and thus the number density for a given exposure), but only
changes the total number of redshifts taken in a single pointing.
Altering the telescope aperture and fibre aperture is a more complex
change as it will affect the exposure times.  This can be accommodated
by changes in the parameters of the number counts calculator.  The
parameters for the different instruments considered are given in Table
\ref{instruments}.

\begin{table}
\begin{center}
\caption{Spectrograph parameters for different instruments. These are
  considered as alternate constraint parameters to those found in
  Table \ref{constraintparameters}. The de-scope option for the WFMOS
  FoV is included in brackets.}
\begin{tabular}{lccc}
\hline
Instrument & WFMOS & AAOmega & SDSS \\
\hline
Overhead time (mins) & 10 & 5 & 30 \\
Min time (mins) & 15 & 60 & 30 \\
$n_{fibres}$ & 3000 & 392 & 640 \\
FoV (diameter in deg) & 1.5 (1) & 2 & 3 \\ 
fibre aperture (arcsec) & 1 & 2 & 3 \\
effective mirror diameter (m) & 8.0 & 3.5 & 2.5 \\
Nod and shuffle & Yes & No & No \\
\hline
\label{instruments}
\end{tabular}
\end{center}
\end{table}

The smaller apertures and larger fibre diameters of the SDSS and
AAOmega systems mean that their exposure times to obtain the same
angular source density as a WFMOS survey are longer, but this is
partially countered by a larger field-of-view which allows them to
survey more of the sky per pointing.  We compared the different
hardware possibilities by finding the best Figure-of-Merit for a
survey with a single bin at low redshift (where all other survey
parameters are allowed to vary). The result of the previous section
showed that the optimal targetted galaxy population changes with the
number of fibres of the instrument, so we also compared targetting red
(continuum) and blue (line emission) galaxies at low redshifts. The
results are shown in Table \ref{instrumentfom}.

\begin{table}
\centering
\caption{Best FoM for surveys conducted on telescopes with the given FoV and Aperture for different galaxy types}
\begin{tabular}{|l|cccc|} \hline
Telescope & Aperture & FoV   & \multicolumn{2}{c}{FoM} \\
 /Instrument & & diameter & blue  & red \\ \hline
Subaru/WFMOS & 8m & 1.5 & 207 & 29 \\
Subaru/WFMOS & 8m & 1.0 & 141 & 16 \\
AAT/AAOMega & 3.5m   & 2   & 13 & 21  \\
SDSS & 2.5m  & 3  &  18 & 21 \\\hline
\end{tabular}
\label{instrumentfom}
\end{table}

We find that line-emitting galaxies are strongly preferred as targets
for WFMOS (regardless of field-of-view), whereas red galaxies are
marginally preferred for SDSS and AAOmega, which only have a few
hundred fibres. This is consistent with the result for WFMOS for a
small number of fibres (see Fig. \ref{fibres}), where red galaxies are
preferred.  We find that the SDSS and AAT are almost equivalent in
terms of Figure-of-Merit, as the smaller aperture of the SDSS system
(and so longer exposure times) is balanced by the larger field-of-view
and number of fibres.  Considering the de-scope option of reducing the
WFMOS field-of-view to 1 degree diameter, this reduces the survey area
by a factor of 2, and thus the FoM drops by almost the same factor (in
detail, since the number of fibres is held constant, the fibre density
increases, which offsets the loss of area to some extent).  In other
words, equivalent results will be obtained by a 3-year survey with a
1.5-deg-diameter system and a 4.5-year survey with a 1-deg-diameter
system.

\subsection{Other Surveys}

During the construction period of a WFMOS-like instrument, a number of
other dark energy surveys will be performed.  These measurements of
the angular diameter distance and Hubble parameter in the case of BAO
surveys, or luminosity distance in the case of Type Ia Supernovae
(SN-Ia) surveys, will have already constrained some of the dark energy
parameter space. By including the predictions for these surveys in our
analysis, we can determine whether our optimal survey design changes.
We consider two surveys that are currently underway: (1) WiggleZ, a
baryon acoustic oscillation survey being carried out with the AAOmega
spectrograph; and (2) the full five-year SuperNovae Legacy Survey and
the Sloan Digitized Sky Survey II Supernova Survey (SNLS-SDSS).

The WiggleZ survey has the following parameters: Area = 1000 sq deg,
$z = 0.75$, $dz = 0.25$, number density $= 8.5 \times 10^{-4}$ $h^3$
Mpc$^{-3}$, observing line-emission galaxies (Glazebrook et al. 2007). Using our
fitting formula code (Blake et al. 2006) we find that this corresponds
to measurement accuracies of 3.1\% in $d_A$ and 5.2\% in $H(z)$. This
is easy to include in our optimization, as it counts as an extra
measurement at a redshift of 0.75, without any time cost.

\begin{figure}
\center
\epsfig{file=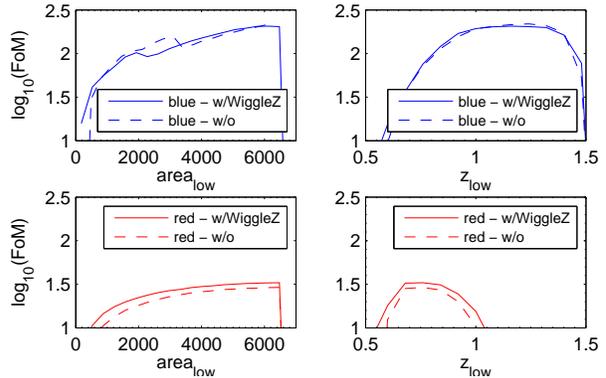, width=8cm}
\caption[WiggleZ]{\label{WFMOS-WiggleZ} The Figure-of-Merit as a
  function of area (left) and redshift (right) for a single-bin survey
  at low redshift using either line emission (top) or continuum
  emission (bottom) galaxies, with (solid) and without (dashed) the
  WiggleZ measurement at $z=0.75$.}
\end{figure}

We find that the inclusion of the WiggleZ measurement has no effect on
any of the survey parameters, as shown in Figure \ref{WFMOS-WiggleZ},
as the curves with and without the WiggleZ survey are virtually
identical.  This is because the WFMOS measurement would produce a much
more accurate measurement of dark energy properties.

The SNLS-SDSS SN-Ia survey will find around 1000 Supernovae
distributed approximately evenly in the range $0.1 < z < 1$ (D. Andrew
Howell and the SNLS collaboration 2004; SDSS-II Supernovae Survey Fall
2005). The measurement is the apparent magnitude ($m$) of the
supernovae, defined as
\beq
m(z) = 5\log_{10}d_L(z)+[M +25] \,.
\eeq

\noindent The luminosity distance will give us information about the
expansion, but we have the added complication of marginalizing over
the absolute magnitude $M$. The supernova surveys therefore do not
constrain the Hubble parameter $h$. The Fisher matrix for supernova
surveys is (Tegmark et al, 1998; Kim et al, 2004)
\beq
F_{ij} = \frac{1}{\sigma_m^2} \int dz \, N(z) \frac{\partial
  m}{\partial p_i} \frac{\partial m}{\partial p_j} \,,
\eeq

\noindent where $\sigma_m$ is the error in the magnitude and $N(z)$ is
the number of SN-Ia in a redshift bin around $z$.  We assume 100 SN-Ia
in each of ten equally-spaced redshift bins from 0.1 to 1, setting
$\sigma_m = 0.12$. Appendix A details the evaluation of the Fisher
matrix elements.

\begin{figure}
\center
\epsfig{file=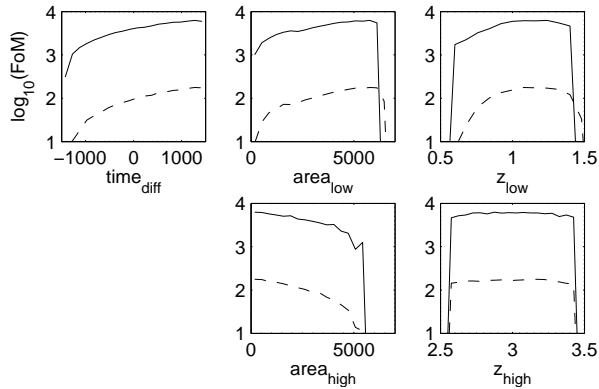, width=8cm}
\caption[snia]{\label{wfmos-snia} The Figure-of-Merit as a function of
  the survey parameters for a two-bin survey using line emission
  galaxies, with (solid) and without (dashed) the SNLS-SDSS SN-Ia
  measurement over the range $z=0.1 - 1$. Although the inclusion of
  the SN-Ia measurement greatly improves the FoM, it has no effect on
  the distribution of survey parameters with Figure-of-Merit, and does
  not change the parameters of the optimal survey.}
\end{figure}

The SN-Ia surveys and the WFMOS BAO survey are approximately
equivalent in their effectiveness at investigating the dark energy,
with the SN-Ia FoM being 158, compared to the WFMOS FoM of 207. When
the two surveys are combined, we find once again that the inclusion of
the SN-Ia measurement has no effect on the distribution of survey
parameters with Figure-of-Merit, as show in Figure
\ref{wfmos-snia}. However, the increase in the Figure-of-Merit is
marked, improving from $\sim207$ to $\sim 6250$, equivalent to a
shrinkage of error ellipse area of about $\sqrt{6250/207} \simeq 5.5$
times. This is due to the complimentary nature of the two types of
measurement.

\section{Conclusions}
\label{conclusions}

In this paper we have conducted an optimization of the survey
parameters of a WFMOS-like survey of the baryon acoustic oscillations
in order to best constrain the concordance $\Lambda$CDM model. We use the determinant of
the $w_0,w_a$ Fisher matrix of the predicted errors on the dark energy
as our Figure-of-Merit to choose between different surveys. We find
that when optimizing for a cosmological constant ($\Lambda$) model of
the dark energy ($w_0=-1$, $w_a=0$), a line-emission galaxy selection
strategy is preferred over continuum emission. The optimal survey
spends all the available survey time (1500 hours in this analysis)
observing at low redshift, surveying an area of around 6000
sq.\ degs. The medium redshift of the bin is about 1.1 and it will
stretch from $z \sim 0.8 - 1.4$.  The optimal solutions are not highly
peaked, and the flexibility bars are often considerable.  Reducing the
area by several hundred sq.\ degs and increasing the target density
(for a fixed observing time) does not change the dark energy
measurements by a significant amount.

We find that the measurement of the baryon acoustic oscillations in a
high redshift ($z=3$) bin is not optimal for testing a $\Lambda$CDM
model.  This is due to the long exposure times for obtaining redshifts
of LBGs compared to $z=1$ galaxies.  In addition, the effect of the
dark energy on the expansion of the universe is negligible at high
redshift (assuming a cosmological constant) and so measuring the BAOs
at this redshift (in addition to low redshift) makes only (at best) a
small extra contribution to the Figure-of-Merit. If we consider other
more exotic dark energy models with non-negligible energy density at
high redshift, it may become more important and favoured by the IPSO
method. Alternatively, the high-redshift BAOs could be measured with a
Lyman-$\alpha$ survey ``piggy-backed'' on the low-redshift galaxy
survey (McDonald \& Eisenstein, 2006).  However the high redshift part
of the survey only serves to cross-check the paradigm and will not
improve statistical error bars on the cosmological parameters. Even
so, the high redshift bin shouldn't be ruled out purely on this basis,
as it may be responsible for completely new discoveries.

We also examine the optimal instrument design, analyzing the
scientific benefit/penalties of different choices. In particular,
increasing the number of fibres has a positive effect on the FoM, as
more objects can be observed simultaneously, and a higher number
density can be achieved for a given exposure time. This FoM increase
continues until the number of fibres reaches some optimal value, at
which point enough galaxies have been observed to easily beat shot
noise effects, and the FoM stops increasing. This optimal value is
around 10{,}000 fibres for line-emission strategies, and 750 for continuum
emission.  The large difference between the two strategies is due to
the ability of the line-emission surveys to reach higher redshifts.
We note that since 10{,}000 fibres only
obtains successful redshifts for about 6000 galaxies, compared
to most efficient number of fibres by this criterion is 2000.  This
survey configuration has a significantly lower FoM compared to the
optimal survey (151, compared to 328 for 10{,}000 fibres), and so the
optimum would be a value around 3000-4000 fibres.
Note that if we were to include the effect of reconstructing the power
spectra on small scales (Eisenstein et al. 2006b), this may increase
the required number of fibres. Reconstruction increases the
performance in measuring the BAOs for any number density, but works
especially well at high number densities, and may therefore increase
the optimal number density.

We find that including measurements made by other baryon oscillation
and supernovae surveys in the total error analysis does not change the
optimal survey, indicating that the analysis we do now should remain
valid into the medium future.

Finally, we have demonstrated how the IPSO technique can be applied to
``realistic'' simulations of redshift surveys, including instrumental
limitations such as fibre number, repositioning overheads and galaxy
number counts models. Since the measurement of baryon acoustic
oscillations is one method to measure the dark energy, IPSO could also
be applied to the survey \& instrumental design that use other methods
such as weak lensing, integrated Sachs-Wolfe or cluster number counts,
or even those with other science goals.

\section*{Acknowledgments}

We thank Sam Barden, Arjun Dey, Daniel Eisenstein, Andrew McGrath, and
the rest of WFMOS team A for helpful advice and comments. DP
acknowledges helpful comments from Pia Mukherjee. CB acknowledges
funding from the Izaak Walton Killam Memorial Fund for
Advanced Studies and the Canadian Institute for Theoretical
Astrophysics.  MK is supported by the Swiss NSF. DP is supported by
PPARC.  RCN thanks the EU for support via a Marie Curie Chair.  We
thank Gemini for funding part of this work via the WFMOS feasibility
study.  We acknowledge the use of multiprocessor machines at the ICG,
University of Portsmouth.

\appendix

\section{Fisher matrix formalism and model definitions}

We use the Fisher matrix formalism to compute the errors on the model
parameters $p_A$, given the observational errors on the measured
quantities $X_i$.  Here $p_A$ includes the optimisation target parameters 
$\theta$ as well as other nuisance parameters. The Fisher matrix is the 
curvature at the peak of the likelihood,
\beq
F_{AB} = \frac{\dd^2(-\log\LL)}{\dd p_A\dd p_B} .
\eeq
Its inverse is a local approximation to the covariance matrix and provides
a lower limit for the errors on the model parameters via the Kramer-Rao bound.

We can rewrite the Fisher matrix via the chain rule to depend only on
observational quantities,
\beq
F_{AB} = \sum_{ij} \frac{\dd X_i}{\dd p_A} F_{ij} \frac{\dd X_j}{\dd p_B} .
\eeq
The observational Fisher matrix is taken to be the inverse of the data covariance
matrix, $F_{ij} = C_{ij}^{-1}$. 

In our case the model parameter vector is $\hat{p}=\{\Omega_m,\omega_m \equiv \Omega_m h^2,w_0,w_a\}$
where the two $w_i$ parameters describe the dark energy equation of state
$w$, the ratio of the pressure to density. We use the parametrisation
\beq
w(z) = w_0 + w_a \frac{z}{1+z} .
\eeq
We also assume that the universe is flat and that the influence of radiation is
negligible so that $\Omega_{DE} = 1-\Omega_m$.

From the observations we recover the comoving distance to redshift $z$, $r(z)$
(from the transverse modes) and its derivative $r'(z)$ (from the radial modes).
More precisely, we recover $y(z)\equiv r(z)/s$ and $y'(z)\equiv r'(z)/s$. The
quantity $s$ is the comoving sound horizon at last scattering. By using the fitting
formula described in a previous paper (Blake et.\ al, 2006) we also recover the fractional errors 
$x=\Delta y/y$ and $x'=\Delta y'/y'$ for a given observational setup. We assume that
the errors $x$ and $x'$ are uncorrelated between each other and between redshift bins.
In this case the covariance matrix is diagonal. Expressed in terms of these quantities
the Fisher matrix becomes
\bea
F_{AB} & = & \sum_i \frac{1}{y(z_i)^2} \frac{\dd y(z_i)}{\dd p_A} 
\frac{\dd y(z_i)}{\dd p_B} \frac{1}{x_i^2} \\
 & & + \sum_i \frac{1}{y'(z_i)^2} \frac{\dd y'(z_i)}{\dd p_A} 
\frac{\dd y'(z_i)}{\dd p_B} \frac{1}{x_i^{'2}}
\eea
Here the sums run over the observational bins. Separating $y$ into the contributions
due to $r$ and $s$, and writing $D_A f \equiv \dd \log(f)/\dd p_A$ for the
logarithmic derivative of a function $f$ we can write the above formula as
\bea
F_{AB} & = & \sum_i \frac{\left(D_A r(z_i) - D_A s\right)\left(D_B r(z_i)-D_B s\right)}{x_i^2} \\
& & + \sum_i \frac{\left(D_A r'(z_i) - D_A s\right)\left(D_B r'(z_i)-D_B s\right)}{x_i^{'2}}
\eea
It remains to compute $D_A r$, $D_A r'$ and $D_A s$.

The comoving distance is given by
\beq
r'(z) = \frac{c}{H(z)} \; , \qquad r(z) = \int_0^z r'(x) dx .
\eeq
Since we are dealing only with logarithmic derivatives we find that all
constants drop out, and we set $c=1$ from now on.
For our simplified cosmological model the Hubble parameter is
\beq
H^2(z) = H_0^2 \left(\Om (1+z)^3 + (1-\Om) f(z;w_0,w_a)\right)
\label{eq:hub}
\eeq
The function $f(z;w_0,w_a)$ describes the evolution of the energy
density of the dark energy and can be integrated directly for our
parametrisation of $w(z)$,
\bea
f(z;w_0,w_a) & = & \exp\left\{3 \int_0^z \frac{1+w(x)}{1+x} dx\right\} \\
& = & (1+z)^{3(1+w_0+w_a)} \times \\ \nonumber
&  & \exp\left\{-3w_a\frac{z}{1+z}\right\}
\eea

At this point we should rewrite the Hubble parameter in terms of our
base parameter set. Specifically we have to replace $H_0^2 = 10^4 \om/\Om$.
Here we can again neglect the factor $10^4$ as it will drop out of the
Fisher matrix computation. Eq.~(\ref{eq:hub}) is now
\beq
h(z)^2 = \om \left( (1+z)^3 + (1/\Om-1)f(z;w_0,w_a)\right) .
\eeq

As the redshift integration for $r(z)$ converges, we know that differentiation
with respect to $p_A$ and integration over $z$ commute. It is therefore 
sufficient to calculate $\dd r'(z)/\dd p_A$. For $r'$ we have that
$\dd_A r'(z) = - \dd_A h(z)/h^2(z)$. Clearly then $D_A r'(z) = -D_A h(z)$.
Of course for $D_A r(z)$ we need to compute
\beq
D_A r(z) = \frac{\int_0^z \dd_A r'(x) dx}{\int_0^z r'(x) dx} .
\eeq
We will now derive explicitely expressions for all $\dd_A r'$.
\bea
\dd_\om r'(z)  & = & - \frac{1}{2 \om h(z)} = -\frac{1}{2\om} r'(z) \\ 
D_\om r'(z) & = & -\frac{1}{2\om}
\eea
(up to an irrelevant constant pre-factor). In this case we can perform
formally the $z$ integration and find 
\beq
D_\om r(z) = -\frac{1}{2\om} = D_\om r'(z) .
\eeq
For the next term we get
\beq
\dd_\Om r'(z) = -\frac{1}{2} \frac{\dd_\Om h(z)^2}{h(z)^3} 
= \frac{1}{2} \frac{\om f(z;w_0,w_a)}{\Om^2 h(z)^3} .
\eeq
There do not seem to be any straightforward simplifications. As always,
$D_\Om r'(z) = h(z) \dd_\Om r'(z)$. The derivative with respect to the
dark energy parameters $w_i$ is just the same as before, except that we
differentiate the $f$ instead,
\bea
\dd_{w_i} r'(z) &=& -\frac{1}{2} \frac{\dd_{w_i} h(z)^2}{h(z)^3} \nonumber \\
&=& - \frac{1}{2} \frac{\om (1/\Om-1) \dd_{w_i} f(z;w_0,w_a)}{h(z)^3}
\eea
where
\bea
\dd_{w_0} f(z;w_0,w_a) &=& 3 \log(1+z) f(z;w_0,w_a) \\
\dd_{w_a} f(z;w_0,w_a) &=& 3 \left( \log(1+z) - \frac{z}{1+z}\right) f(z;w_0,w_a) 
\eea

For the logarithmic derivative of $s$ we use a numerical differentiation
of the following formulae: the sound horizon at recombination is given by
\beq
s = \frac{c}{\sqrt{3}} \frac{1}{H_0 \Omega_m^{1/2}} \int_0^{a_r}
\frac{da}{(a+a_\equ)^{1/2}} \frac{1}{(1+R)^{1/2}} \,,
\eeq
which can be evaluated (Efstathiou and Bond 1999) and gives
\bea
s & = & \frac{4000}{\sqrt{\omega_b}} \frac{\sqrt{a_\equ}}{\sqrt{1+\eta_\nu}} \nonumber \\
& & \ln\left\{ \frac{[1+R(z_r)]^{1/2} +
[R(z_r)+R_\equ]^{1/2}}{1+\sqrt{R_\equ}} \right\}  {\mathrm Mpc}\,,
\eea
where $\eta_\nu=0.6813$ is the ratio of the energy density in neutrinos
to the energy in photons. The parameter $R=3 \rho_b/4\rho_\gamma$ is
numerically
\beq
R(a) = 30496 \omega_b a \,.
\eeq
The scale factor at which radiation and matter have equal density is
\beq
a_\equ^{-1} = 24185 \left(\frac{1.6813}{1+\eta_\nu}\right) \omega_m
\eeq
and the redshift of recombination, $z_r$, is given by the following
fitting-formula (Hu \& Sugyiama 1996)
\bea
z_r &=& 1048 \left(1+0.00124\omega_b^{-0.738}\right)\left(1+g_1
\omega_m^{g_2}\right) \,, \\
g_1 &=& 0.0783 \omega_b^{-0.238} \left(1+39.5 \omega_b^{0.763}
\right)^{-1} \,, \\
g_2 &=& 0.560 \left( 1+21.1 \omega_b^{1.81} \right)^{-1} \,.
\eea

\end{document}